\documentclass[twocolumn,showpacs,preprintnumbers,amsmath,amssymb]{revtex4}

\pdfoutput=1

\usepackage{graphicx}
\usepackage{dcolumn}
\usepackage{bm}
\usepackage{subfigure}

\begin{document}

\preprint{APS/123-QED}

\title{Fermi-edge Singularity in II-VI Semiconductor Resonant Tunneling Structures}



\author{M. R\"{u}th}
\author{T. Slobodskyy}%
\author{C. Gould}
\author{G. Schmidt}
\author{L.W. Molenkamp}
\affiliation{Physikalisches Institut (EP3), Universit\"{a}t W\"{u}rzburg, Am Hubland, D-97074 W\"{u}rzburg, Germany}

\date{\today}

\begin{abstract}
We report on the observation of Fermi edge enhanced resonant tunneling transport in a II-VI semiconductor heterostructure. The resonant transport through a self assembled CdSe quantum dot survives up to 45 K and probes a disordered two dimensional (2D) like emitter which dominates the magnetic field dependence of the transport. An enhancement of the tunnel current through many particle effects is clearly observable, even without an applied magnetic field. Additional fine structure in the tunneling current suggests that while conventional Fermi edge singularity theory successfull reproduces the general features of the increased transmission, it is not adequate to describe all details of the current enhancement.
\end{abstract}

\pacs{Valid PACS appear here}
\maketitle

\begin{figure}[tb]
	\centering
		\includegraphics{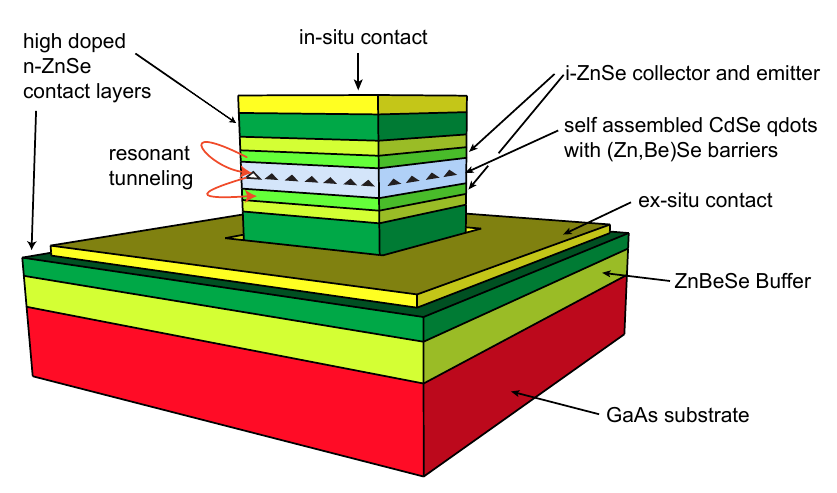}
	\caption{Device layout and layer structure. Self assembled quantum dots which are embedded in the (Zn,Be)Se tunnel barrier provide the resonant state for the tunneling transport mechanism.}
	\label{fig:RTD_big_noMn}
\end{figure}

\begin{figure}[tb]
	\centering
		\includegraphics{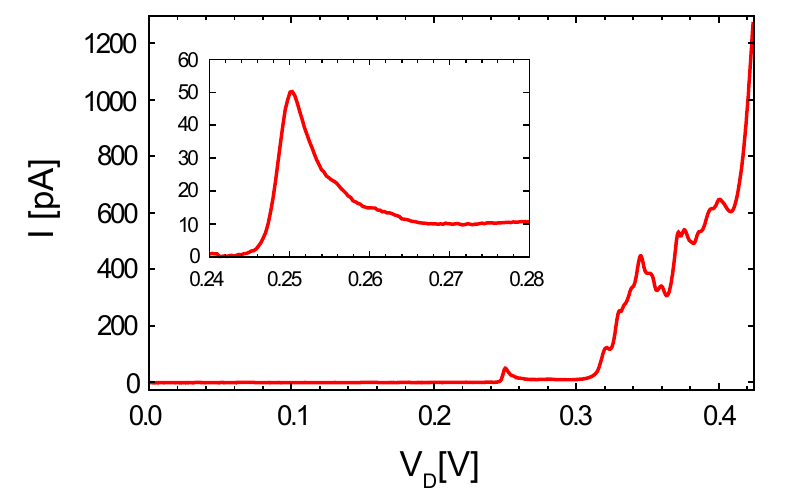}
	\caption{Forward bias I(V$_D$) characteristic at 4.2 K and zero magnetic field. The inset shows a magnification of the first peak, at 250 mV, which has the shape of a Fermi edge singularity enhanced resonance.}
	\label{fig:bto5FullRange}
\end{figure}

\begin{figure}[tb]
	\centering
		\includegraphics{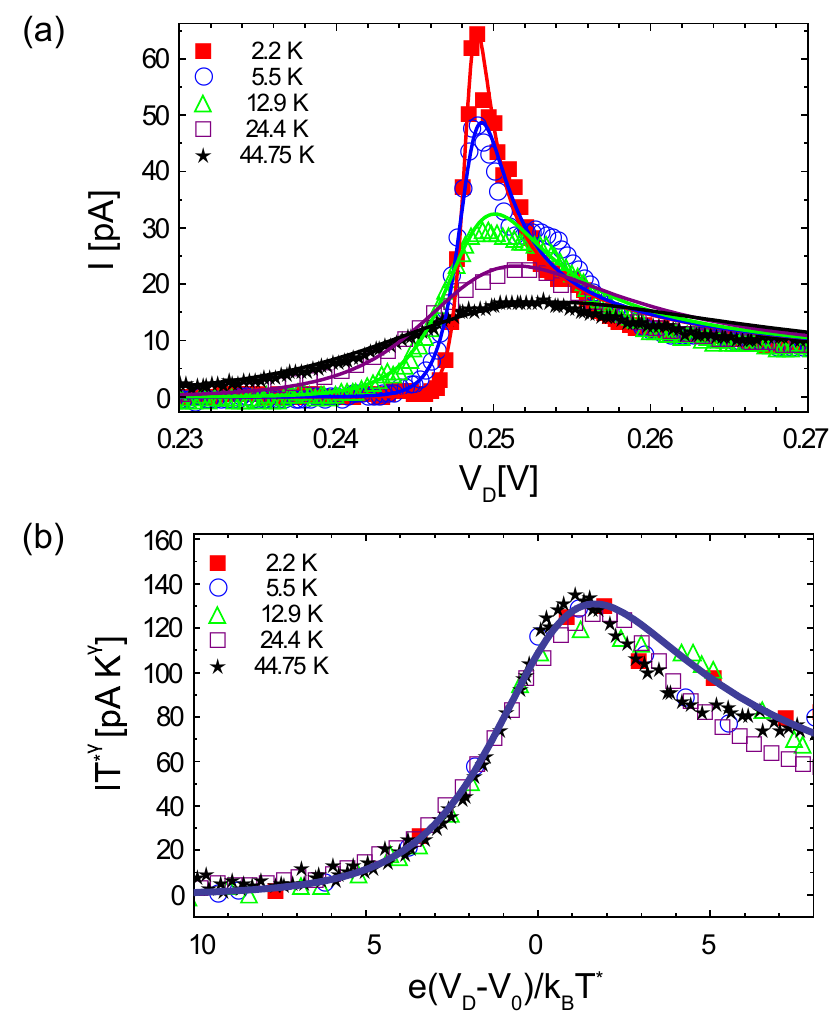}
		\caption{(a) (color online) I(V$_D$) characteristics for various temperatures up to 45 K and at zero magnetic field. Symbols are the experimental data. The colored line for each dataset represents a fit to equation \ref{eq:fermi} for each temperature. (b) Rescaling both axes collapses the datasets for all temperatures on a single scaling curve. The solid line is a fit to the rescaled equation \ref{eq:fermi} which is now independent of the effective temperature $T^*$.}
	\label{fig:together}
\end{figure}

\begin{figure}[tb]
	\centering
		\includegraphics{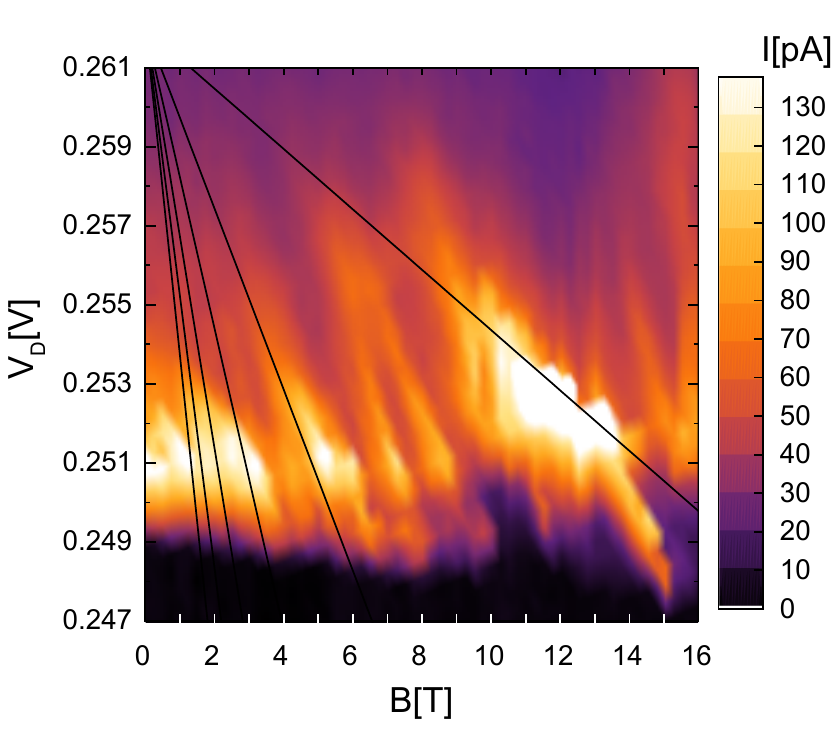}
	\caption{Current as a function of bias voltage and magnetic field perpendicular to the layer stack at 25 mK. Several field dependent resonance features are observed.}
	\label{fig:cb2158_magneticfield1}
\end{figure}

\begin{figure}[b]
	\centering
		\includegraphics{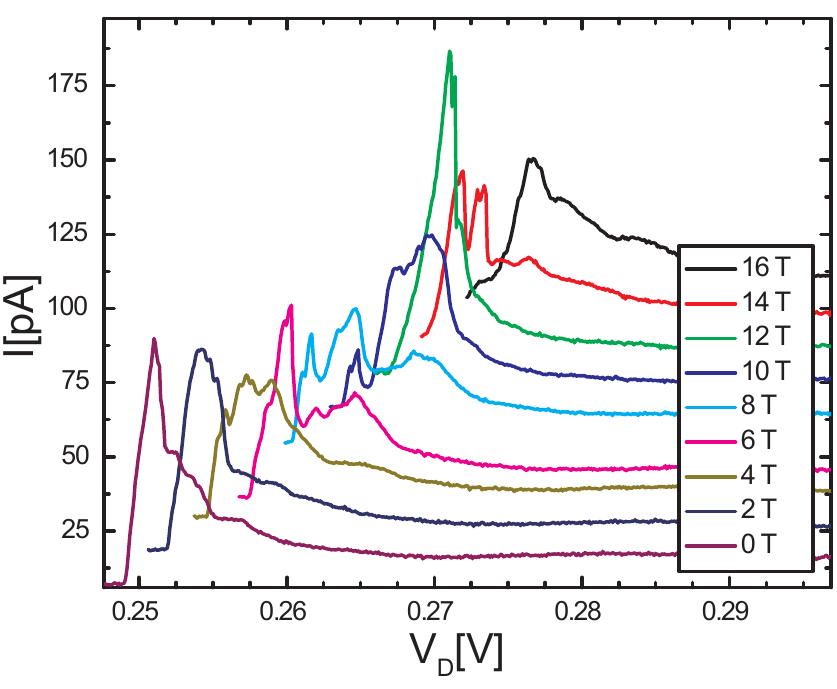}
		\caption{I(V$_D$) characteristics for various magnetic fields at 25 mK. An additional fine structure on the classical Fermi edge singularity can be resolved (The curves are offset by 3 mV and 11 pA for clarity). }
	\label{fig:waterfall}
\end{figure}

Resonant tunneling transport is currently topical for its potential applications in spintronics and future electronics. Moreover, it serves as a useful testbed for investigations of fundamental transport properties. In particular, many body correlation effects can significantly influence the transmission probability in zero dimensional transport through a double barrier heterostructure, and one of the ways in which they manifest themselves are through Fermi edge singularities (FES). Previous experimental studies of FES have been limited to III-V devices, often at high magnetic fields. In this letter we report on the observation of a FES in an all-II-VI semiconductor resonant tunneling diode (RTD). The feature is already clearly observed in the absence of external magnetic field for temperatures up to 45 K, and shows magnetic field dependence reflective of two dimensional (2D) character in the injector.

Our device is a molecular beam epitaxy (MBE) grown all-II-VI semiconductor RTD with an active region consisting of a Zn$_{0.7}$Be$_{0.3}$Se tunnel barrier with embedded self assembled CdSe quantum dots. Because of the lattice mismatch between the CdSe and the barrier material, the growth of 1.3 monolayers of CdSe into the middle of the barrier induces a strain build up. Relaxation of this strain results in the formation of quantum dots. Emitter and collector are gradient n-type doped, with an undoped spacer layer next to the barriers and a high doping concentration ($10^{19}$ cm$^{-3}$) at the contact layer to allow the formation of ohmic contacts at the metal-semiconductor interface. The layer structure and the device layout are shown in fig. \ref{fig:RTD_big_noMn}. In order to allow for transport measurements vertically through the layer stack, standard optical lithography techniques are used to pattern 10 x 10 $\mu m^2$ square pillars, and contacts are applied to the bottom and top ZnSe layers. The contact on top of the RTD pillar is deposited immediately after growth by transfering the sample to a ultra-high-vacuum metallization chamber without breaking the vacuum. It has ohmic behavior and a contact resistance on the order of $10^{-3}\: \Omega cm^2$ as determined by stripline measurements on callibration samples. Bottom contacts must be deposited ex-situ after processing the pillar and thus have higher resistivity which is compensated in our device by the larger area (500$^2$ $\mu$m$^2$) of these contacts. Although for the size of our pillar one expects some ten thousand self assembled quantum dots in each device, transport through self assembled quantum dot RTDs at lower bias voltages is usually dominated by only a few dots that have resonant levels at relatively low energies and/or at weak spots in the barriers \cite{Hill2001,Vdovin2000,Patane2002,Itskevich1996,Hapke-Wurst2000}. Therefore a resonant feature at low bias is characteristic of resonant tunneling from the injector through a single self assembled quantum dot into the collector.

Figure \ref{fig:bto5FullRange} shows the I(V$_D$) characteristic of our sample at 4.2 K in zero magnetic field and under forward bias (the top of the pillar being defined as ground). At bias voltages above 300 mV, the current shows many resonance-peaks superimposed on the normal background. These originate from the ensemble of quantum dots in the structure. The inset shows the first resonant feature of our device at 250 mV which is separated by approximately 60 mV from the next visible resonance. The shape of this feature is clearly suggestive of FES behavior \cite{MATVEEV1992,GEIM1994,Frahm2006}. The FES causes an additional flow of electrons on resonance with the emitter Fermi level. This enhancement decays as a power law with increasing bias voltage \cite{MATVEEV1992}. At 1.6 K, it has a maximum value of seven times the current flowing at higher biases (say 30 mV after the resonance onset), where many particle effects are negligible.

Figure \ref{fig:together}(a) illustrates the temperature dependence of the I-V characteristics. The resonant feature survives to high temperature with the current enhancement decreasing by only a factor of 4.7 from 2 to 45 K. As observed before in III-V devices \cite{Vdovin2007}, the maximum current of the FES enhanced resonance decays with increased temperature according to a power law. We fit the results using the temperature dependent modelling developed by Frahm et al. for a FES enhanced resonant tunneling current through a self assembled InAs quantum dot \cite{Frahm2006}. The tunneling current is given by

\begin{eqnarray}
	&I(V_D,T^*)\\
&\propto\nonumber\frac{1}{\pi}\mbox{Im}\left[\left(\frac{i D}{\pi k_{B}T^*}\right)^\gamma B\left(\frac{1-\gamma}{2}-i\frac{\alpha(V_D-V_0) e}{2 \pi k_{B}T^*},\gamma\right)\right]
	\label{eq:fermi}
\end{eqnarray}

with a lever arm $\alpha =0.4$, the characteristic cutoff parameter $D$ of the order of the bandwidth and the edge-exponent $\gamma$. $B$ represents the Beta-function. An effective Temperature $k_BT^*=\sqrt{(kT)^2+\Gamma_i^2}$ is used to account for the effect of the intrinsic linewidth $\Gamma_i$ of the participating quantum dot state. As shown experimentally in fig. \ref{fig:together}(b), and consistent with equation \eqref{eq:fermi}, the I(V$_D$) curves taken at various temperatures collapse to a common curve after rescaling voltage and current axes with $V_D\rightarrow e(V_D-V_0)/k_BT^*$ and $I\rightarrow IT{^*}^\gamma$, where the Fermi-edge exponent $\gamma=0.54$ and the intrinsic bandwith $\Gamma_i=0.25$ meV are used as fitting parameters. The model agrees with experiment, with only a slight deviation at higher bias voltages, which probably stems from a small thermal drift during measurement.

The current of a FES enhanced tunneling process is influenced by magnetic field \cite{Frahm2006,Vdovin2007,Hapke-Wurst2000}. Figure \ref{fig:cb2158_magneticfield1} shows transport measurements at 25 mK in various magnetic fields perpendicular to the layer stack, which clearly shows Landau fan like structure.

In general, this structure could result from electronic states either in the emitter or in the dot. If it was related to the quantum dot, well known Darwin-Fock behavior would lead to an increase of the energy of the resonant Landau levels with increasing magnetic field, which is in stark contrast to the observations of fig. \ref{fig:cb2158_magneticfield1}. Thus we conclude that the magnetic field dependence of all features is dominated by the emitter states, which gain energy with magnetic field and thus reduce the peak position of each feature.

The black lines in fig. \ref{fig:cb2158_magneticfield1} represent an emitter Landau fan, calculated assuming a bulk ZnSe effective mass $m\approx 0.18 m_e$ and the same lever arm determined by the previous fit at zero magnetic field. Many important features are fitted by the Landau fan, however some additional field dependent features that are present cannot be explained by using only this one Landau fan or by including the Darwin-Fock like behavior of the quantum dot level.

Resonant tunneling transport showing magnetic properties of the emitter layer, where all main features could be explained by one Landau fan, has been observed in a III-V device with self assembled InAs quantum dots and a high quality two dimensional electron gas (2DEG) as emitter \cite{Main1998}. We suggest that in our device we measure a 2D-like disordered emitter which results in various additional resonance signals in the energy-magnetic field plane due to a more complicated density of state structure in the emitter. These additional features cannot be attributed to a second dot or an additional dot level. In that case, one would expect two resonant features for each emitter Landau level, separated by an almost constant voltage offset. It is however not possible to match the data with two such Landau fans. Moreover a second dot should show a clearly visible feature at zero magnetic field, but none is observed (see fig. \ref{fig:bto5FullRange}). Direct evidence of the 2D characteristics of the emitter is that the Landau structure is only visible for the perpendicular magnetic field configuration, and vanishes for measurement in in-plane magnetic field.

While the model of equation \eqref{eq:fermi} correctly describes the shape of the Fermi edge singularity at temperatures above 1 K, measurements at 25 mK (fig. \ref{fig:cb2158_magneticfield1}) reveal additional fine structure which is not explained within this model. To emphasize the nature of this fine structure, I(V$_D$) characteristics for some of the measured magnetic fields from fig. \ref{fig:cb2158_magneticfield1} are plotted in fig. \ref{fig:waterfall}. This fine structure may result from properties of the emitter. A generalized approach to the FES problem shows that backscattering in the contacts may have an effect on the FES exponent \cite{Abanin2004}. In our case, back scattering within the disordered emitter is a function of the applied bias voltage and produces additional structure on the FES enhanced tunneling current. This could result from characteristics of the local geometry near the dot or local density fluctuations in the disordered-metal-like emitter \cite{Schmidt2001}.

In summary, we have observed a Fermi edge singularity in tunneling from an intrinsic ZnSe injector through a self assembled CdSe quantum dot state. Above 1 K and at zero magnetic field, the behavior of the Fermi edge enhancement is consistent with previous observations in III-V devices, and consistent with conventional theory. At lower temperatures or in a magnetic field, additional enhancement effects are observed, and are suggestive of backscattering associated with local effects near the dot. This detailed characterization of Fermi edge singularity in the II-VI material system thus provides supplementary data against which modern theories of Fermi edge enhanced transport can be tested.

Acknowledgement: The authors acknowledge financial support from the
German Schwerpunktprogramm SPP1285.

\end{document}